# Correlations among physical properties of pervious concrete with different aggregate sizes and mix proportions

Qifeng Lyu[*], Pengfei Dai, Anguo Chen

School of Urban Construction, Changzhou University, Changzhou, Jiangsu 213164, China

**Abstract**: Permeable pavement material can benefit urban environment. Here in this work, different aggregate sizes and mix proportions were used to manufacture pervious pavement concrete and investigate correlations among its properties. The porosity, permeability, compressive strength, inner structure, thermal conductivity, and abrasion resistance of the specimens were obtained. Results showed lower aggregate-to-cement ratios and higher water-to-cement ratios led to porosity reduction, which decreased the permeability coefficient but increased the compressive strength, thermal conductivity, and abrasion resistance of the pervious concrete. Compared to the mixes, the aggregate sizes affected the physical properties of pervious concrete less. However, the sizes of pores and cement in the pervious concrete were more affected by aggregate sizes than by mixes. Moreover, the porosity, permeability coefficient, and compressive strength of the pervious concrete can be correlated by the power law, whereas the correlation between the porosity and abrasion resistance index can be fitted by a linear law.

*: lqf@cczu.edu.cn

# 1 Introduction

Pervious concrete is a kind of pavement material that can be used in light-traffic roads, parking lots, and footpaths (Cao et al., 2022; Hammodat, 2021; C. Huang et al., 2023; Z. Li et al., 2022), although traditional pavement is usually impermeable to provide high mechanical strengths and durability. The reason for using the pervious concrete in pavements is to alleviate relevant environmental problems brought by the traditional impermeable pavement. For example, impermeable pavement may cause water-logging on rainy days (Elizondo-Martínez et al., 2020; Güneyisi et al., 2016; Liu et al., 2023), that is, rainwater cannot be drained off in time and then floods the city street and buildings (Kayhanian et al., 2012). Furthermore, impermeable pavement may also cause urban heat island effect on sunny days (Elizondo-Martínez et al., 2020; Lima et al., 2022), due to the high thermal conductivity and low heat capacity of impermeable pavement (Beddaa et al., 2023; Seifeddine et al., 2022). On hot summer days, this effect may raise the urban temperature by more than 6°C. Moreover, traffic noise on impermeable pavement is also a serious problem (Elizondo-Martínez et al., 2020; Lou et al., 2022). To alleviate such problems, pervious concrete is suggested to be used in the construction of urban infrastructures, especially for light-load pavement which may not require much higher mechanical strengths, but the environmental benefits are more important (Park et al., 2021; Rodrigues et al., 2022; A. Singh et al., 2022).

Pervious concrete is usually made of single-graded coarse aggregates, a small amount of cement, and water (Aliabdo et al., 2018; Lima et al., 2022; Seifeddine et al., 2023), which has the merits of increasing rainwater infiltration, decreasing surface runoff (Seifeddine et al., 2023; Vieira et al., 2020), improving water quality (Wu et al., 2022), adjusting environment temperature (Beddaa et al., 2023; Güneyisi et al., 2016), absorbing traffic noises (El-Hassan et al., 2019), etc. To fully take advantage of these functions, many studies had been done (Elizondo-Martínez et al., 2020; Seifeddine et al., 2023). Generally, the porosity, permeability coefficient, and compressive strength of the pervious concrete are 15%~35%, 0.2~1.2 cm/s, and 2.8~28 MPa (Aliabdo et al., 2018; Wei Huang & Wang, 2022; Lima et al., 2022), respectively. It is true that the high porosity increases permeability but decreases strength and durability. This is an optimization problem in the production of pervious-concrete. To improve the performance of pervious concrete, many influence factors had been studied, including aggregate properties (Fan et al., 2023; Jike et al., 2022; Vieira et al., 2020), cement content (Chaitanya & Ramakrishna, 2021; Wei Huang & Wang, 2022; Zheng et al., 2023), admixture types (da Silva et al., 2022; Lima et al., 2022; Mehrabi et al., 2021), and compaction methods (Güneyisi et al., 2016; Strieder et al., 2022). Additionally, recycled aggregate can be used in the manufacture of pervious concrete (Basavana Gowda et al., 2023; El-Hassan et al., 2019; Zhu & Jiang, 2023), which is not only economic and sustainable but also improves the porosity and permeability (Güneyisi et al., 2016; Strieder et al., 2022; Yap et al., 2018). However, recycled aggregate can also decrease the mechanical strength of the pervious concrete (El-Hassan et al., 2019; Sriravindrarajah et al., 2012; Strieder et al., 2022). Moreover, adding fibers can increase the flexural and tensile strengths and abrasion resistance of the pervious concrete (L. G. Li et al., 2021; Mehrabi et al., 2021; S. B. Singh & Madasamy, 2022). But fibers can also weaken the coated cement, especially when the fiber content is higher. From literature review, it can be found that the macroscopic mix proportions associated with the mesoscopic structures of the pervious concrete greatly influence the comprehensive performances of the pervious concrete, and thus further research on mix proportions and structures of the pervious concrete should be carried out.

To investigate the structures of the pervious concrete, many methods exist. Porosity test can give a macroscopic view for the pores fraction in the pervious concrete. Thanks to the relatively large pores, the porosity of the pervious concrete can be directly measured by the water infiltration method (Cai et al., 2021; Liu et al., 2023; Taheri et al., 2021), which is akin to mercury intrusion porosimetry (MIP) but MIP is more suitable for the micropores in cement (Ke et al., 2021). In addition to MIP, gas absorption (Brunauer-Emmett-Teller, BET method) can also measure the micropores and the pore size distributions (Tang et al., 2020). Clearly, the three-dimensional (3D) information on the macropores usually governs the physical properties of the pervious concrete. The magnetic resonance imaging (MRI) method (Hafid et al., 2015; Tang et al., 2020) and X-ray computed tomography (X-CT) method (Deng et al., 2023; Lyu, Dai, et al., 2023; Yang et al., 2019) can both provide 3D information on pores in the pervious concrete. However, MRI needs to trace the hydrogen in water (Quan et al., 2022), and thus merely give information on permeable pores. For isolated pores, cement, and aggregate, X-CT can provide detailed 3D reconstruction (Cepuritis et al., 2017; Lyu, Chen, et al., 2023; Lyu et al., 2021). However, the densities of cement and aggregate are close to each other. This would bring tremendous difficulties in the segmentation of cement and aggregate on the CT images (Deng et al., 2023; Fan et al., 2023; Lyu, Dai, et al., 2023). Therefore, new advances in segmentation methods should be researched.

To investigate the influences and correlations of physical properties of the pervious concrete as a pavement material, this work used different aggregate sizes and mix proportions to manufacture 20 groups of pervious concrete. The porosities, permeability coefficients, compressive strengths, thermal conductivities, and abrasion resistance indices of the specimens were obtained and analyzed. A high-energy X-CT facility was used to detect the inner structures of the pervious concrete, and a manual watershed segmentation method was used to segment the pores, cement, and aggregates. New data and insights on the correlations of physical properties of the pervious concrete were obtained, which may provide insights and knowledge for better or intelligent design of permeable pavement in the future.

## 2 Materials and Methods

### 2.1 Materials

Natural gravels were used for the coarse aggregates in the experiment. The properties of the aggregates are shown in Table 1 which contains particle size, bulk density, and apparent density. The aggregates were divided into four groups according to their sizes, which were named A1 (4~6 mm), A2 (6~9 mm), A3 (9~12 mm), and A4 (12~15 mm), respectively, and a single gradation was used in the manufacturing of pervious concrete. The average bulk density and average apparent density of the aggregates are 1654.35 kg/m$^3$ and 2797.78 kg/m$^3$, respectively, and slight deviations exist in different groups. The water absorption of the aggregates is 0.45%.

Table 1. Properties of aggregates.

| Aggregate | A1 | A2 | A3 | A4 |
| --- | --- | --- | --- | --- |
| Particle size (mm) | 4~6 | 6~9 | 9~12 | 12~15 |
| Bulk density (kg/m$^3$) | 1655.1 | 1654.5 | 1654.4 | 1645.4 |
| Apparent density (kg/m$^3$) | 2752.4 | 2745.5 | 2872.3 | 2820.9 |

For the cementitious material, grade P·O 42.5 (Tianshan brand) was used, which was purchased from Wuxi, China. The bulk density, apparent density, and fineness of the cement are 1600 kg/m$^3$, 3150 kg/m$^3$, and 350 m$^2$/kg, respectively. The particle size distribution of the cement was tested by laser granularity analysis and the result is shown in Fig. 1a, where it can be found the cement particle sizes are generally distributed between 0.001 and 0.08 mm. The crystal phases of the cement and the hydrated cement past were tested by X-ray diffraction (XRD). Results are shown in Fig. 1b. The phases of unhydrated cement consist mainly of calcium silicate ($3CaO \cdot SiO_2$ and $2CaO \cdot SiO_2$), calcium aluminate ($3CaO \cdot Al_2O_3$), calcium ferro-aluminate ($4CaO \cdot Al_2O_3 \cdot Fe_2O_3$), and calcium sulfate ($CaSO_4$). After hydration, the hardened cement past mainly consists of portlandite, ettringite, calcite, quartz, gypsum, and also a small amount of unhydrated calcium silicate.

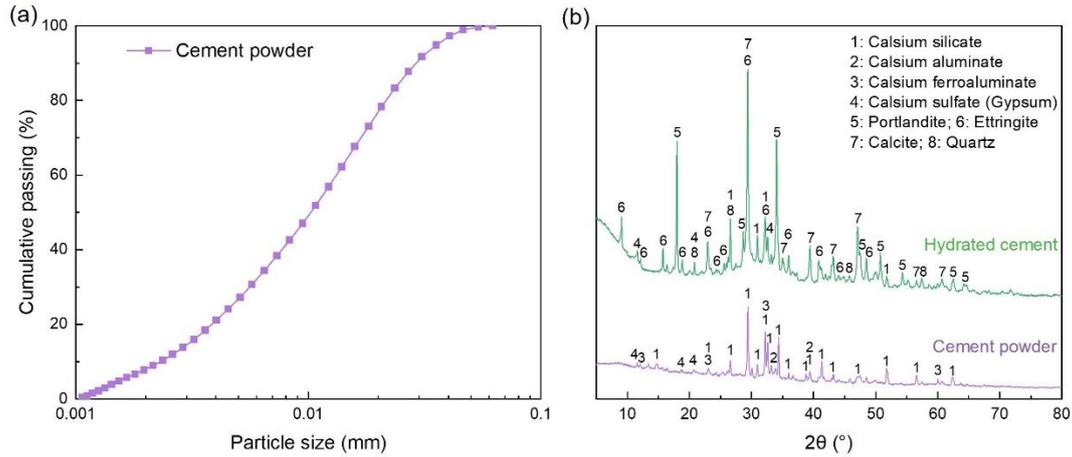

Fig. 1. (a) Particle size distribution and (b) XRD patterns of the cement.

The aggregates and cement were mixed with tap water according to the proportions listed in Table 2 to manufacture the pervious concrete. In the experiments, the aggregate-to-cement ratio varied among 3, 3.5, and 4, whereas the water-to-cement ratios of 0.25, 0.275, and 0.3 were used. Additionally, the aggregates were grouped by size as listed in Table 1. Thus, 20 groups of pervious concrete were finally manufactured. The group names are shown in Table 3. In each group, three specimens were taken to perform physical properties tests.

Table 2. Mix proportion of the pervious concrete by mass ratio.

| Mix | Aggregate | Cement | Water |
|-----|-----------|--------|-------|
| M1  | 3         | 1      | 0.3   |
| M2  | 3.5       | 1      | 0.3   |
| M3  | 4         | 1      | 0.3   |
| M4  | 4         | 1      | 0.275 |
| M5  | 4         | 1      | 0.25  |

Table 3. Specimen groups of the pervious concrete.

| Group | A1 | A2 | A3 | A4 |
|-------|------|------|------|------|
| M1 | M1A1 | M1A2 | M1A3 | M1A4 |
| M2 | M2A1 | M2A2 | M2A3 | M2A4 |
| M3 | M3A1 | M3A2 | M3A3 | M3A4 |
| M4 | M4A1 | M4A2 | M4A3 | M4A4 |
| M5 | M5A1 | M5A2 | M5A3 | M5A4 |

A horizontal concrete mixer was used to mix the raw materials. At the beginning of the mixing procedure, the aggregate and cement were added to the mixer stirring for one minute. Subsequently, the water was gradually poured into the mixture, and the stirring continued for another two minutes. After the mixing, the mixture was cast into 100 × 100 × 100 mm cubic mold. A metal rod was used to compact the mixture (fresh pervious concrete) by tamping the mixture surface 20 times at every half height of the mold during the casting. The specimens were demolded 24 hours later, and then cured in a room with constant temperature and humidity (20±2°C, 95%+) for 28 days.

**2.2 Properties test methods**

The porosity of the pervious concrete was tested according to ASTM C1754/C1754M-12 (Cai et al., 2021; Taheri et al., 2021). The test method is shown in Figs. 2a and 2d, where the masses of the specimen in dry and wet states were measured and noted as $m_d$ and $m_w$, respectively. The specimen's volume is represented by $V$, and the density of water is $\rho$. Then the porosity of the specimen can be calculated by

$$\varphi = \left(1 - \frac{m_d - m_w}{\rho V}\right) \times 100\% \tag{1}$$

A constant head permeameter was used to measure the permeability coefficient of the pervious concrete. The test method and facility are shown in Figs. 2b and 2e, which are in accordance with JC/T2558-2020 based on the Darcy theory that relates the fluid velocity in the pervious concrete with constant pressure drop of the fluid (Lv et al., 2014). The permeability coefficient $K$ can be calculated by

$$K = \frac{QL}{A\Delta h}, \tag{2}$$

where $Q$ represents the volumetric flow rate of the outlet, $L$ represents the specimen's length along the fluid flow direction, $A$ represents the cross area of the specimen, $\Delta h$ represents the pressure drop of the fluid.

The compressive strength of the specimens was tested in accordance with GB/T50081-2019 (Wu et al., 2022). A hydraulic servo universal test machine was used in the test, cf. Figs. 2c and 2f. The size of the test specimen is 100 × 100 × 100 mm, and the loading speed was set at 0.5 MPa/s. Three specimens in each group were tested, and the average and standard deviation of the three tests were calculated and recorded.

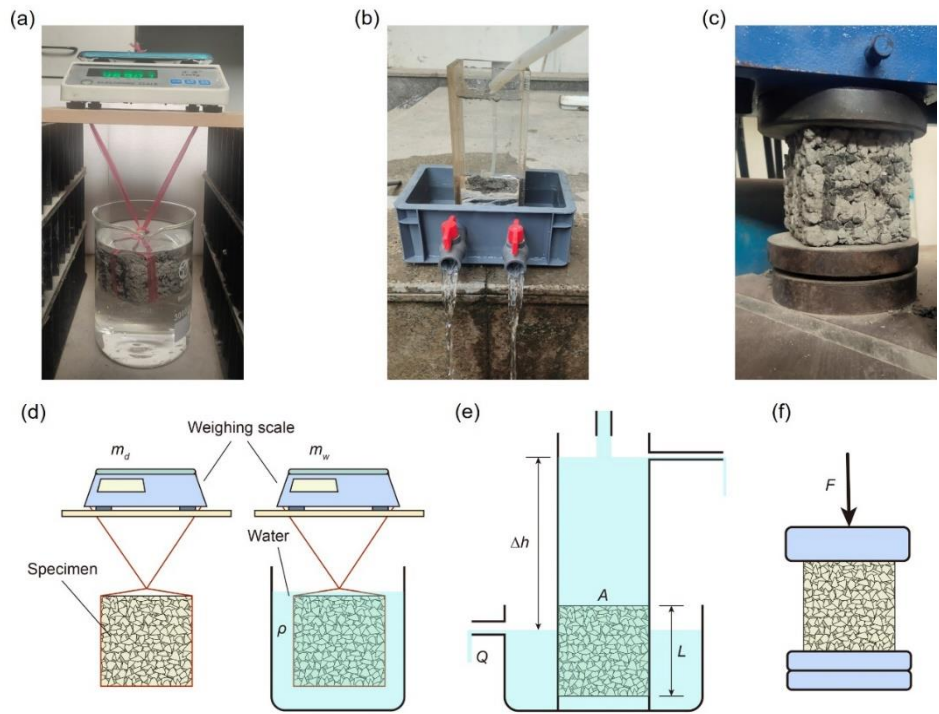

Fig. 2. Test methods for (a, d) porosity, (b, e) permeability, and (c, f) compressive strength of the pervious concrete.

**2.3 Inner structure test methods**

A high-energy industrial X-CT was used to detect the inner structure of the pervious concrete as shown in Fig. 3. The facility mainly consists of three parts that are X-ray source, specimen table, and X-ray receiver. During a test, the X rays were generated from the source, then penetrated the pervious concrete on the specimen table, and last reached the X-ray receiver with attenuated intensities which reflect the varying densities of the tested specimen. As the specimens had a relatively large size of 100 × 100 × 100 mm, the test voltage and current were set at 450 kV and 3.3 mA, respectively, to generate powerful X rays that can fully penetrate the specimen (Fan et al., 2023). The data were stored as 32-bit float images on disk with 2048 × 2048 pixels in each slice and a resolution of 0.098 mm/pixel.

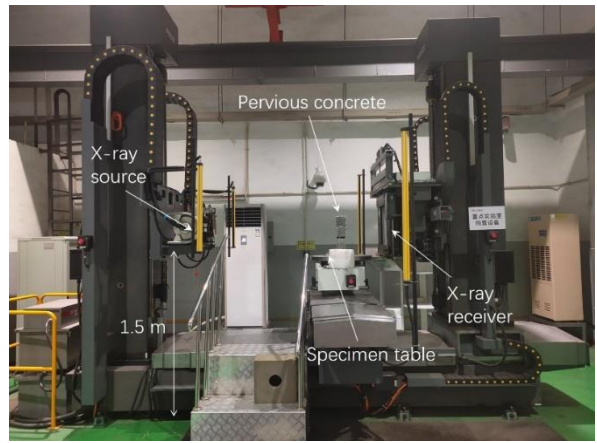

Fig. 3. Industrial X-ray CT.

The CT images were reconstructed in VG Studio and then exported as raw data to be analyzed in Avizo, where the median filter was used to denoise the data, and then a watershed segmentation method was used to differentiate the pores in the pervious concrete. However, segmentation of cement and aggregate is difficult because their densities are close to each other and the specimen sizes are relatively large. A method of manually selecting the watershed seed area was used, which used the lasso tool in the watershed segmentation editor to tag the affirmed cement and aggregate pixels on the CT slice, and then the tagged seed areas could be expanded to the entire range of the materials according to the watershed segmentation algorithm. After the segmentation of pore, cement, and aggregate, the fraction of the components can be calculated.

**2.4 Thermal conductivity calculation methods**

Based on the CT structures, the thermal conductivity of the pervious concrete can be calculated, which is a crucial factor affecting the temperature adjustment and durability of the pavement (Bayraktarova et al., 2023), and also helps to understand the temperature load when the concrete is served as pavement in all-weather conditions. Therefore, the reconstructed CT structures were used to calculate the thermal conductivity of the pervious concrete, as shown in Fig. 4. During the calculation, a 1 K temperature difference was applied on the upper and lower surfaces of the CT reconstructed pervious concrete which is deemed a two-component material. Each component of the concrete has a different thermal conductivity: 1.28 W/m·K and 0.024 W/m·K for the solid structure and air, respectively. Then the following equations were solved:

$$\nabla^2 T = 0, \quad \text{in } \Omega, \tag{3}$$

$$\mathbf{H} = -\tau_i \nabla T, \quad \text{in } \Omega_i, \tag{4}$$

where $\nabla^2 = \partial^2/\partial x^2 + \partial^2/\partial y^2 + \partial^2/\partial z^2$ is the Laplacian operator; $T$ is the temperature; $\mathbf{H}$ (vector) is the heat flux; $\tau_i$ is the thermal conductivity of the $i$th component; $\nabla = \partial/\partial \mathbf{x}$ is the vector differential operator, in which $\mathbf{x}$ represents the coordinates vector in the space, i.e., $x$, $y$, $z$; $\Omega$ is the whole calculated domain; $\Omega_i$ is the $i$th component domain ($i = 1, 2$ for the structure and air, respectively), $\Omega_1 \cup \Omega_2 = \Omega$ and $\Omega_1 \cap \Omega_2 = \emptyset$. The inlet and outlet of the modeling domain were connected with heat reservoirs at different constant temperatures. Then the heat flows under the temperature gradient. A method of mesh-free finite element with linear basis functions was used to solve these equations. After the solving, the thermal conductivity of the specimen can be calculated by

$$\tau = \frac{HL}{V \mathrm{d}T}, \tag{5}$$

where $H$ is the total heat flux, $L$ is the specimen's length along the heat flow direction, $V$ is the volume of the specimen, and $\mathrm{d}T$ is the temperature drop between the inlet and outlet of the specimen.

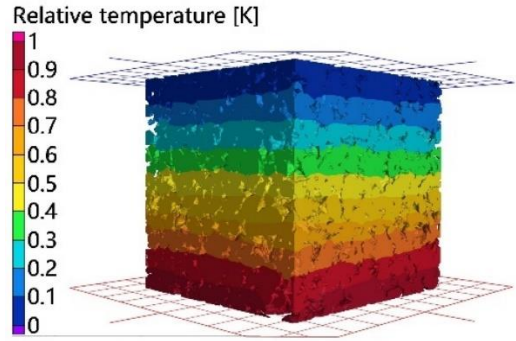

Fig. 4 Thermal conductivity calculation methods

**2.5 Abrasion resistance model**

The abrasion resistance ability of pervious concrete is a crucial factor in evaluating the durability of the pavement material under long-term dynamic load. Several models existed to correlate the abrasion resistance with the compressive strength (H. Li et al., 2006). Here a model in Ref. (Raj & Chockalingam, 2020) is used to calculate the abrasion resistance index $I$ of the previous concrete, which can be written as

$$I = 2.62 \ln F_c - 0.3, \tag{6}$$

where $F_c$ represents the compressive strength (MPa) of the specimen. The abrasion resistance index $I$ is a dimensionless indicator to evaluate the weight loss of the pavement material during the Cantabro abrasion test (Wentong Huang et al., 2022; Raj & Chockalingam, 2020). It is defined by the following equation:

$$I = \frac{\sqrt{R/R_B}}{W_L/W_I}, \tag{7}$$

where $R$ represents the number of revolutions with a speed of 30 rpm in a Cantabro abrasion test, $R_B$ represents the benchmark number of revolutions (300) at 30 rpm, $W_L$ is the weight loss in the Cantabro abrasion test (kg), $W_I$ is the initial weight of the specimen (kg). Therefore, a higher abrasion resistance index indicates a lower weight loss in the Cantabro abrasion test. That is to say, the specimen has good durability in resistance to abrasion.

## 3 Results and discussion

### 3.1 Porosity

Porosities of the specimens are shown in Fig. 5. The maximum porosity is 31.1% belonging to a specimen in group M5A2, whereas the minimum porosity appears in group M1A3 with a value of 6.87%. Furthermore, from Fig. 5, it can be found that three height platforms for the porosities grouped by mix proportions exist. Specifically, M1 is the first level with an average porosity of 10.2%. M2, M3, and M4 are on the second level with their average porosities of 20.2%, 21.2%, and 22.6%, respectively, which are indeed close to each other. M5 has the highest porosities, with an average value of 29.1%. This trend of change exists for different aggregate sizes along the mix coordinate, indicating the mix proportions affect the porosity more than the aggregate sizes do.

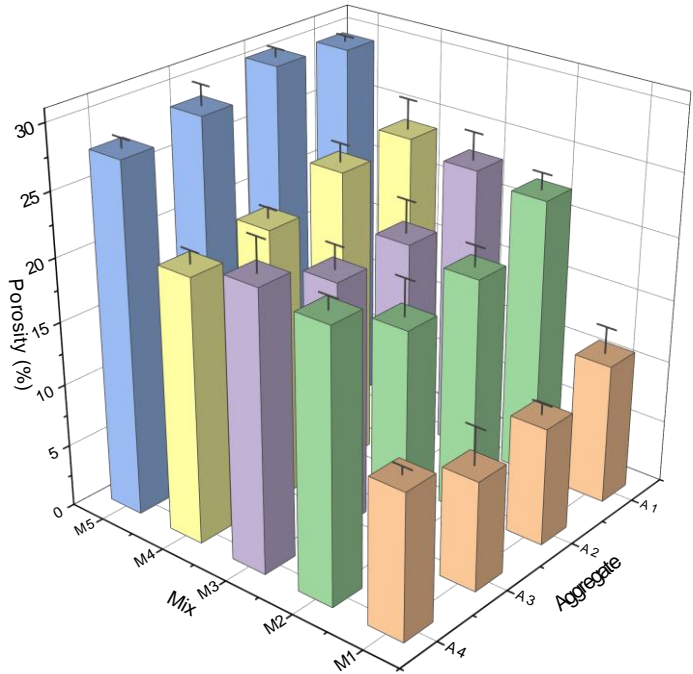

Fig. 5. Porosities of the pervious concrete.

Considering the mix proportions listed in Table 2, M3 can be selected as the focus to be compared and analyzed, since it occupies the center of not only the mix designs but also the porosity measurement results. Specifically, M3 has the aggregate-cement-water ratio of 4:1:0.3. Reducing the aggregate-to-cement ratio to 3.5:1 (M2), as in the middle column of the violin plot shown in Fig. 6a, affects the porosity little. However, when the aggregate-to-cement ratio decreases to 3:1 (M1), a distinct decrease appears in porosity. In fact, the average porosity has decreased by 51.9% from M3 to M1. Similarly, when reducing the water-to-cement ratio from 0.3 (M3) to 0.275 (M4), the porosity increases slightly (6.6%). But when continuing to reduce the water-to-cement ratio to 0.25 (M5), cf. Fig. 6b, the porosity increases by 37.3% which is much greater than the former 6.6%. However, the aggregate size affects the porosity less. This can be inferred from Fig. 6c, where the average porosities for specimens in groups A1, A2, A3, and A4 are 22.1%, 20.5%, 19.4%, and 20.8%, respectively, which show a V trend with slight deviations. But the spans of the porosity in each group are large, which all cover from 5% to 35%. By grouping the data in each aggregate-size group according to their mix proportions, five curves connecting the average porosities in each group are shown in Fig. 6d. Specifically, M1, M2, and M3 show similar V trends as that shown in Fig. 6c, whereas M4 and M5 show downward trend with a slight deviation for the data point of M5A2. The differences are mainly due to the randomness in the manufacturing process. This also indicates the influence of aggregate size on the porosity is less than that of mix proportions.

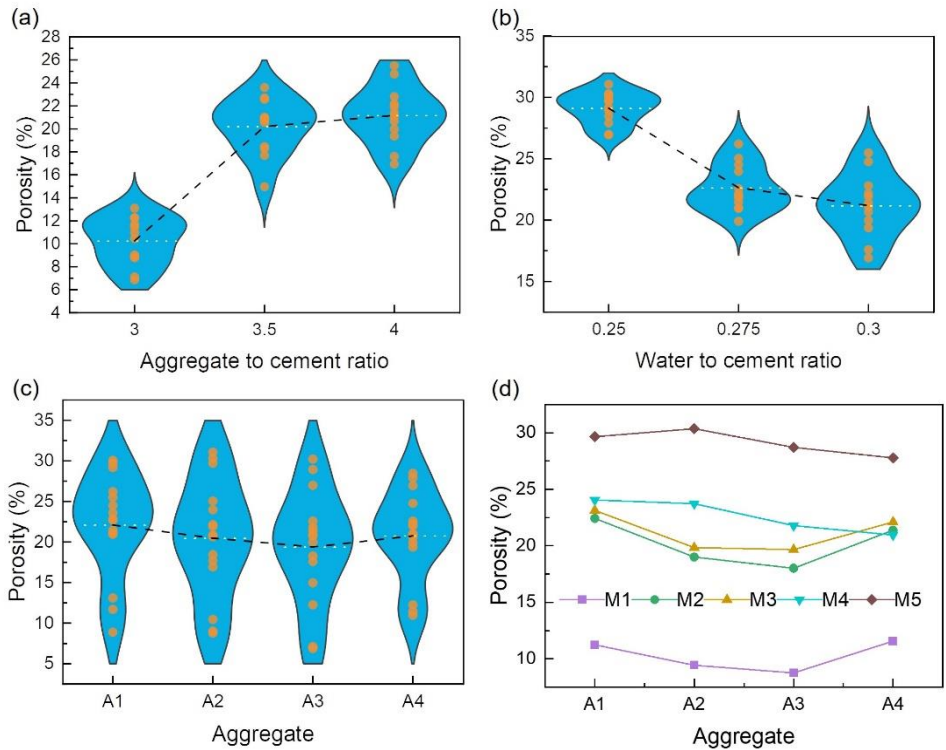

Fig. 6. Influence of (a) aggregate-to-cement ratio, (b) water-to-cement ratio, (c) aggregate size, and (d) mix proportion on the porosities of pervious concrete.

From the above results, it can be concluded that the porosity of the pervious concrete mainly depends on the mix proportions. And smaller aggregate-to-cement ratios and larger water-to-cement ratios may lead to lower porosities. The reason is the cement paste in these situations is apt to flow and may block the pores among the aggregates. This phenomenon is shown in Fig. 7. It can be found that the top surfaces of the fresh pervious concrete in groups M2A2 (Fig. 7a) and M5A1 (Fig. 7b) show different glosses. The higher gloss in Fig. 7a indicates the water-to-cement ratio in group M2 is greater. Although the top surface of M2A2 lacks much cement past in comparison with that of M5A1 in Fig.7b because the metal roller took the surface cement away during the rolling compacting process, the bottom surface of M2A2 in Fig. 7c shows a large amount of deposited cement paste due to the driving forces of gravity and compacting. This is also contributed by the smaller aggregate-to-cement ratio of M2, as the cement content is relatively high in this situation. Generally, the deposited cement paste would block the pores like those shown in the yellow ellipses. This reduces the porosity and may subsequently affect the permeability and strength of the pervious concrete.

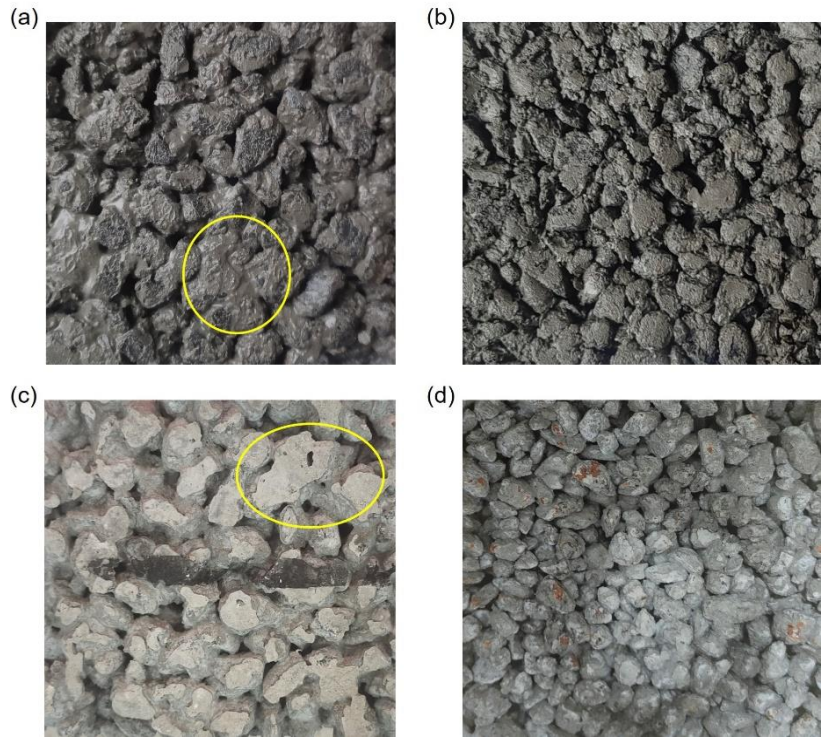

Fig. 7. Top surfaces of specimens in groups (a) M2A2 and (b) M5A1 during the manufacturing process; Bottom surfaces of specimens in groups (c) M2A2 and (d) M5A1 after the curing.

**3.2 Permeability**

The permeability coefficients of the specimens are shown in Fig. 8, where it can be found similar three height platforms of the permeability coefficients grouped by mix proportions exist, although some fluctuations blur the trend, especially the specimens in group M2A4. Specifically, the average permeability coefficient of specimens with mix M1 is 0.128 cm/s which is the lowest. M2, M3, and M4 occupy the middle platform, whose average permeability coefficients are 0.808 cm/s, 0.711 cm/s, and 0.670 cm/s, respectively, whereas the highest average permeability coefficient 1.165 cm/s appears in mix M5. Generally, the change trends of permeability coefficients are similar to those of porosities. Also, this indicates the mix proportions affect the permeability coefficients more than the aggregate sizes do. On the other hand, the fluctuations in the permeability coefficients indicate the permeability coefficients do not merely depend on porosity. Some other factors, such as aggregate size, tortuosity, and mix proportion also affect the permeability coefficient.

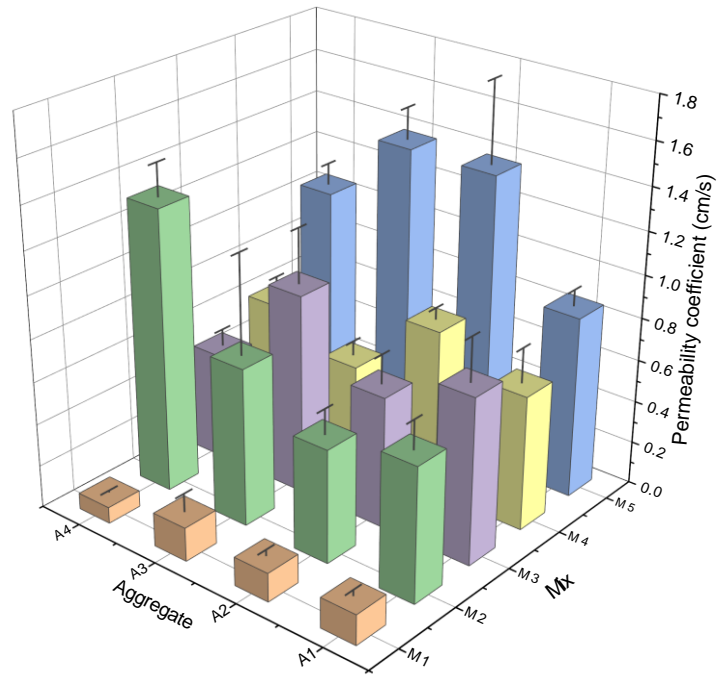

Fig. 8. Permeability coefficients of the pervious concrete.

The influences of mix proportions and aggregate sizes on the permeability coefficients are shown in Fig. 9. For the aggregate-to-cement ratio, when it decreases from 4 (M3) to 3 (M1), the permeability coefficient generally decreases too, although the changes in M2 and M3 are trivial, cf. Fig. 9a. Furthermore, the distribution range of the data points in M1 is smaller. This is because the redundant cement paste blocked the pores and then reduced the permeability coefficients and their deviations. Similarly, the water-to-cement ratio affected the permeability coefficient as it affected the porosity. A lower water-to-cement ratio generally leads to a higher permeability coefficient, although the influence of 0.275 ~ 0.3 is relatively small, cf. Fig. 9b. When comparing the data grouped by aggregate sizes, as shown in Fig. 9c, no obvious influence appears in the average permeability coefficients. However, after continuing to group the data according to the mix proportions, the aggregate size shows influences on the average permeability coefficients of the specimens, cf. Fig. 9d. But the influences are rather random, and no uniform change trend can be concluded for the curves. This is because the randomness of the specimens and their manufacturing cause the permeability coefficients to be highly complex.

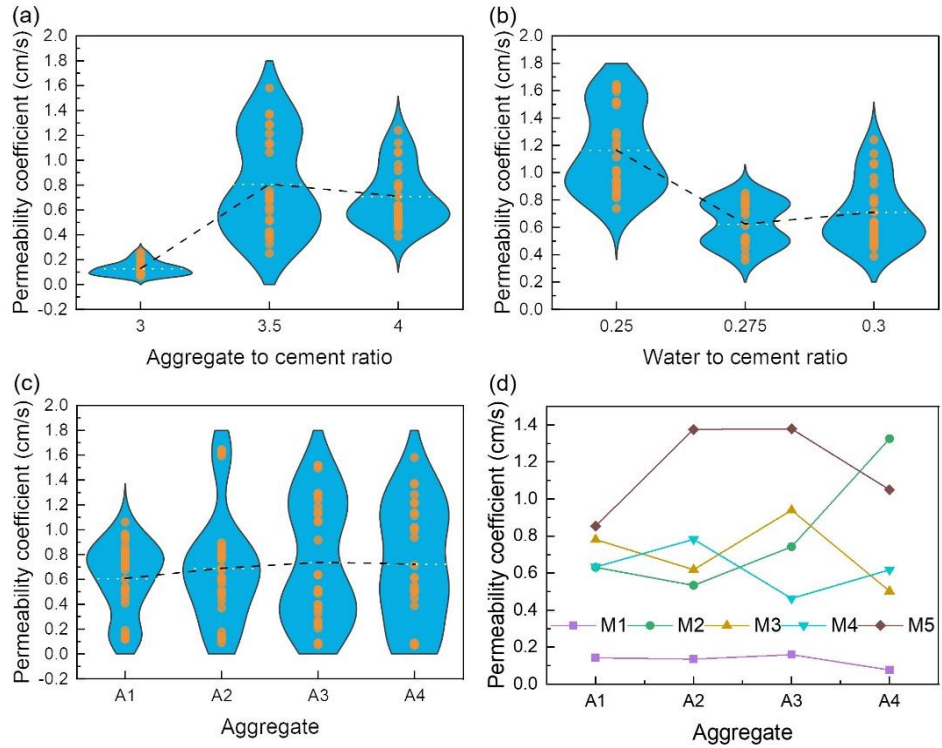

Fig. 9. Influence of (a) aggregate-to-cement ratio, (b) water-to-cement ratio, (c) aggregate size, and (d) mix proportion on the permeability coefficients of pervious concrete.

**3.3 Compressive strength**

The compressive strengths of the specimens are shown in Fig. 10. Unlike the porosities and permeability coefficients in Figs. 5 and 8, the compressive strengths of the specimens show no obvious platforms grouped by the mix. Instead, M1 has the highest compressive strengths, whereas those in the other mixes decrease significantly. Specifically, the average compressive strength of M1 is 18.33 MPa, and those of M2, M3, M4, and M5 are 9.925 MPa, 10.03 MPa, 7.448 MPa, and 6.290 MPa, respectively. The maximum compressive strength of all specimens is 24.68 MPa which occurred in a specimen of M1A1, whereas the minimum compressive strength is 3.69 MPa that belongs to M5A2. Additionally, the aggregate size also affects the compressive strength, although the influence is not significant in comparison with that of the mix proportion. Details can be found in Fig. 11.

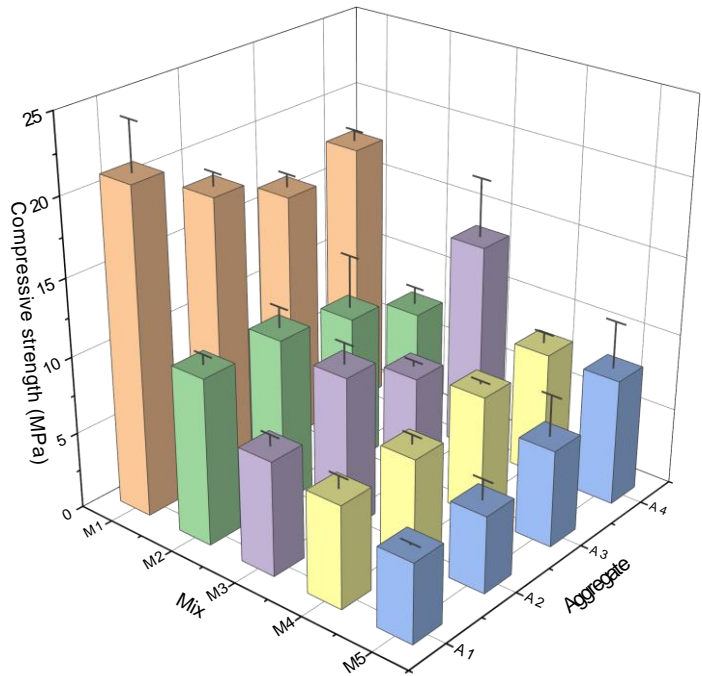

Fig. 10. Compressive strengths of the pervious concrete.

Similar to porosity and permeability, the compressive strength is also affected by the aggregate-to-cement ratio. This is depicted in Fig. 11a, where it can be found when reducing the aggregate-to-cement ratio from 4 to 3.5, the average compressive strength of the specimens changes little. However, when continuing to reduce the aggregate-to-cement ratio to 3, significant increase in the compressive strength appears. This is due to the redundant cement paste that blocked the pores and subsequently increased the compressive strength. The reason is similar, but the change trend is contrary to that of porosity and permeability of the specimens. For the water-to-cement ratio, it also shows opposite change trend to that of porosity and permeability, though the change of compressive strength between 0.275 and 0.3 (water-to-cement ratio) is more significant, cf. Fig. 11b. This indicates the water-to-cement ratio affects the compressive strength more directly than the aggregate-to-cement ratio dose. The aggregate size also affects the compressive strength but not significantly. This can be found in Fig. 11c. Similar as before, after continuing to group the data by mix proportions, the influence of the aggregate size on the compressive strength appears. However, the change trends of the curves are not consistent, cf. Fig. 11d. Specifically, M1 and M2 show a downward trend when increasing the aggregate size, whereas M4 and M5 show an upward trend, and M3 shows a zigzag trend. The reason is also similar to that of permeability, namely that the compressive strength mainly depends on the mix proportion of cement paste rather than the aggregate size which generally contributes more randomness to the physical properties of the pervious concrete.

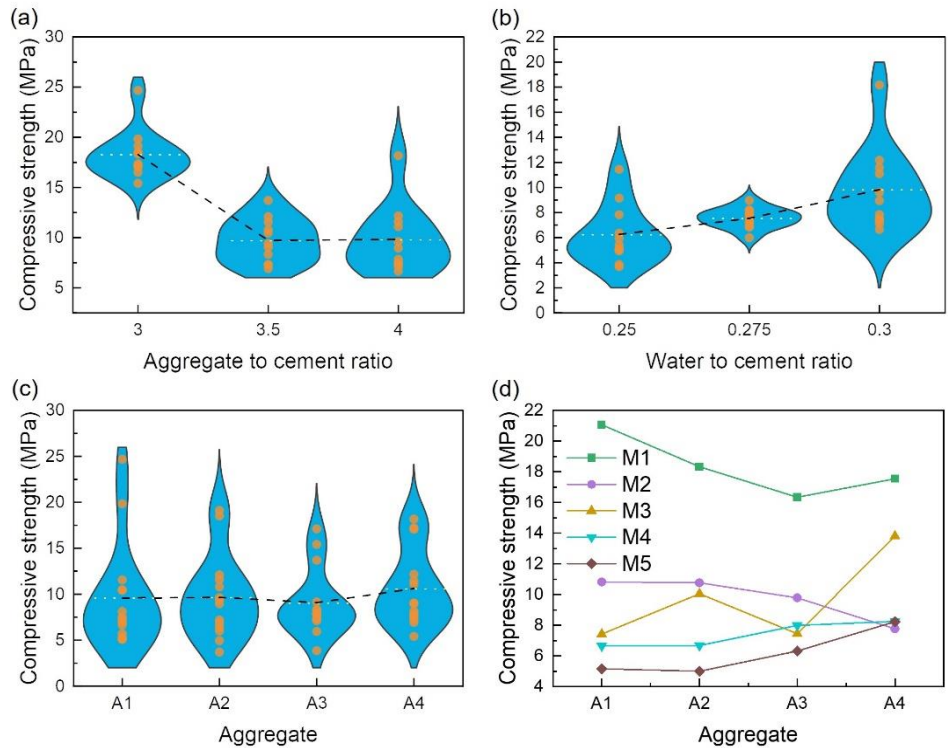

Fig. 11. Influence of (a) aggregate-to-cement ratio, (b) water-to-cement ratio, (c) aggregate size, and (d) mix proportion on the compressive strengths of pervious concrete.

The failure characteristics of the specimens after the compressive strength test are shown in Fig. 12. Generally, two types of failure appeared in the specimens. The first is the failure of aggregate, for example, those in the yellow ellipses in Fig. 12. This type of failure usually occurred in high-strength specimens with relatively plenty of cement paste and large aggregate size. The other type of failure is due to the failure of cement paste, like those in the red ellipses in Figs. 12b and 12c. This type of failure is more abundant in the pervious concrete, since the high porosity may cause the cracks to be generated along the aggregate boundary which is the weak point under pressure. To enhance the mechanical strengths, this kind of failure (failure of cement paste) should be reduced as much as possible.

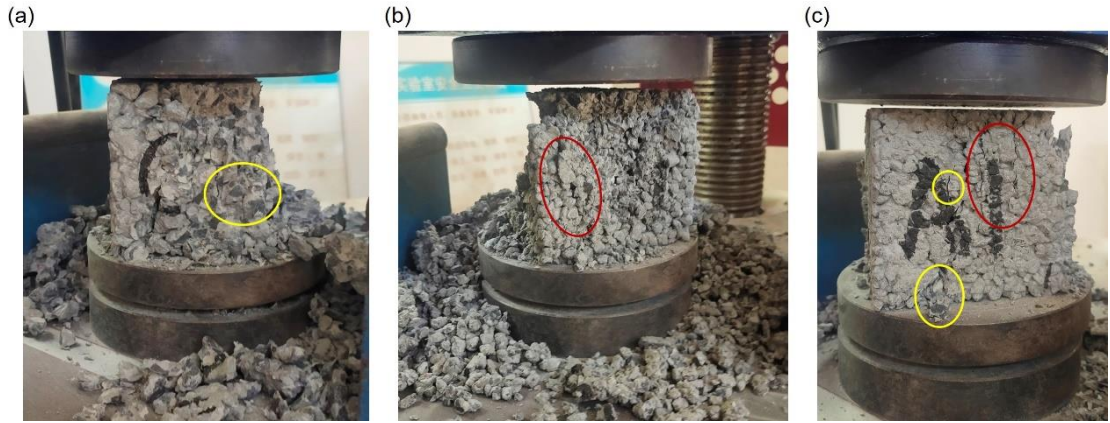

Fig. 12. Failure of specimens in groups (a) M3A2, (b) M5A1 and (c) M1A1.

**3.4 Pores and cement distribution**

Pores in the pervious concrete can be segmented from the X-CT images using the watershed segmentation method. Results are shown in Fig.12 which illustrates the pore distribution of specimens in groups M1A3, M2A3, M3A3, M5A3, M4A1, M4A2, M4A3, and M4A4. The upper-row subfigures in Fig.13 compare the specimens with the same aggregate size but different mixes. It can be found the mixes affect the pore distribution greatly. In Fig. 13a, the majority pores of M1A3 are isolated, whereas pores in other specimens of Figs. 13b, 13c, and 13d are denser and well-connected. The porosities of the specimens in Figs 13a, 13b, 13c, and 13d measured by the X-CT method are 5.7%, 8.1%, 14.5%, and 19.3%, respectively, which are all lower than those measured by the experiment shown in Fig.2. This is due to the truncation error that the micropores below 0.098 mm cannot be detected by the X-CT in order to provide a large-size scan. The lower-row subfigures in Fig. 13 depict how the aggregate size affects the pore distributions. Generally, the influence exists mainly in the pore sizes which increase with the aggregate size. The porosities are also affected, but not significantly. For the specimens shown in Figs. 13e, 13f, 13g, and 13h, whose porosities are 21.4%, 18.5%, 15.1%, and 16.2%, respectively. Additionally, the pore distributions are not uniform in the specimens, especially for those with high water-to-cement ratio and low aggregate-to-cement ratio, like M1, M2, and M3, because the cement paste in such specimens is apt to flow under gravity or compacting forces, which may randomly block some pores.

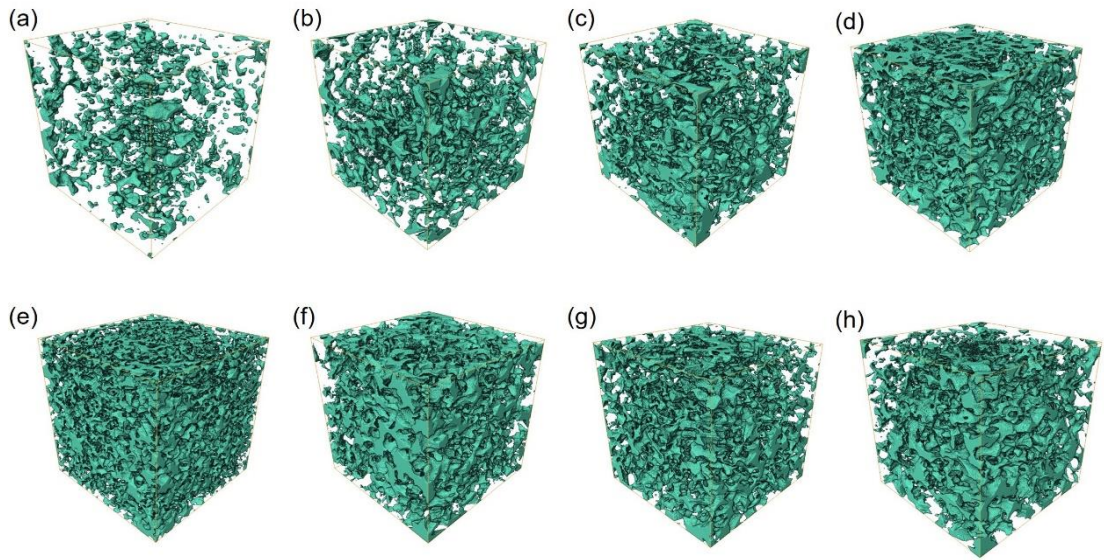

Fig. 13. Pore distributions of specimens in groups (a) M1A3, (b) M2A3, (c) M3A3, (d) M5A3, (e) M4A1, (f) M4A2, (g) M4A3 and (h) M4A4.

The cement distributions, together with the pores and aggregates, are shown in Fig. 14, although segmentation of the three materials is difficult. The method shown in Section 2 is better but slightly time-and-labor intensive. Therefore, only five slices of each specimen were segmented and Fig. 14 shows the typical ones. The fractions of the pores, cement, and aggregates were calculated for the five segmented slices. Results are shown in Fig. 15. For specimens of M1A3, M2A3, M3A3, M4A3, and M5A3 whose aggregate sizes are the same, their pore fractions show an upward trend, whereas the cement paste fractions show a V trend and the aggregate fractions show a zigzag trend. However, for specimens with the same mix but different aggregate sizes, i.e., M4A1, M4A2, M4A3, and M4A4, their fractions of pore, cement, and aggregate merely show trivial fluctuations as those shown in Fig. 15. This indicates the aggregate sizes affect the fractions less than the mixes do.

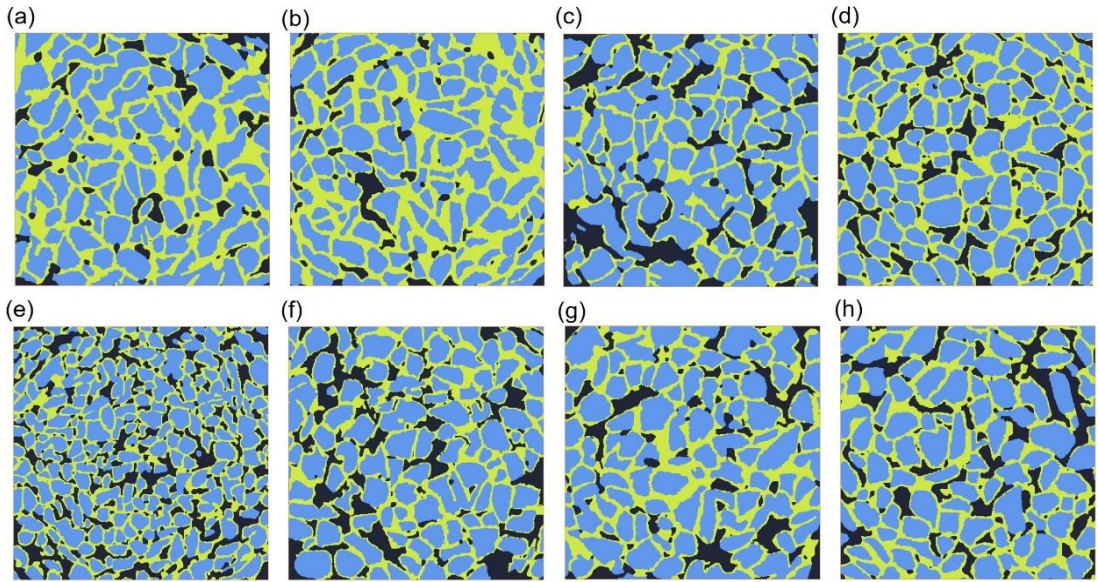

Fig. 14. Pore, cement, and aggregate distributions on slices of specimens in groups (a) M1A3, (b) M2A3, (c) M3A3, (d) M5A3, (e) M4A1, (f) M4A2, (g) M4A3, and (h) M4A4.

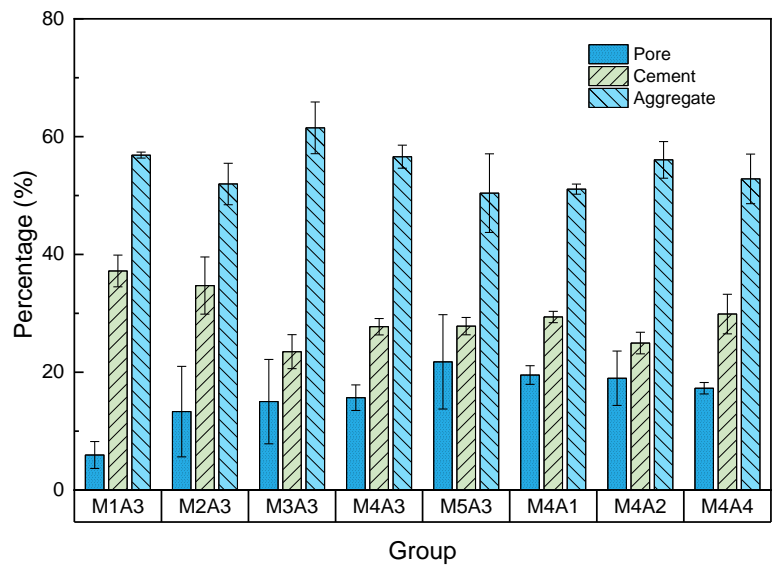

Fig. 15. Pore, cement, aggregate fractions of specimens in groups M1A3, M2A3, M3A3, M4A3, M5A3, M4A1, M4A2, and M4A4.

### 3.5 Thermal conductivity

The thermal conductivity together with the volumetric porosity of the aforementioned CT reconstructed specimens are shown in Fig. 16, where it can be found the thermal conductivity shows a V trend in the specimen groups, whereas the porosity shows a totally reverse trend, indicating the thermal conductivity correlates with the porosity significantly. The reason for the significant correlation is the pores in the pervious concrete, which determine the porosities directly, can reduce the thermal conductivity because the air in the pores has extremely lower thermal conductivity. On the whole, the thermal conductivities of all calculated specimens cover a span of 0.85-1.21 W/m·K which is lower than that of the impermeable solid structure (1.28 W/m·K), and also in a reasonable range compared with those in the literature (Beddaa et al., 2023; Seifeddine et al., 2022). For the porosities, they are relatively lower than those measured in experiments, because the micropores whose sizes are below the resolution were omitted in the CT segmentation and reconstruction (Lyu, Dai, et al., 2023), although the trend is reasonable to explain how the thermal conductivity changes in the specimens.

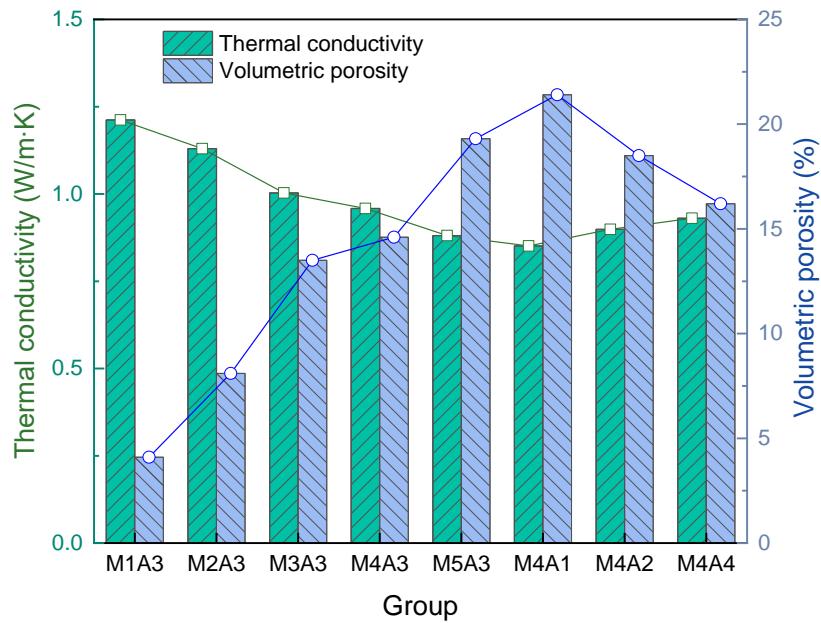

Fig. 16. Thermal conductivity and volumetric porosity of the CT reconstructed specimens in groups M1A3, M2A3, M3A3, M4A3, M5A3, M4A1, M4A2, M4A4.

**3.6 Abrasion resistance**

The average abrasion resistance indices of the pervious concrete in each group are shown in Fig. 17. Generally, the indices cover a range of 1.83-4.51, which is similar to that measured in the literature (Raj & Chockalingam, 2020). From Fig. 17, it can be also found that the mix M1 has relatively higher abrasion resistance indices, indicating such specimens are good at resisting the abrasion. The reason is that the cement content used in this mix is higher. For the specimens in other mixes, their abrasion resistance indices decrease gradually along the mix sequence number, although those in M3 have higher fluctuations. On the whole, the average abrasion resistance indices of M1, M2, M3, M4, and M5 are 4.61, 2.95, 2.86, 2.23, and 1.72, respectively. Similar to the compressive strength, the maximum abrasion resistance index of all specimens is 5.4 which appears in a specimen of M1A1, whereas the minimum abrasion resistance index is 0.42 that belongs to M5A2. Additionally, the aggregate size also affects the abrasion resistance index, though the influence is not significant in comparison with that of the mix proportion.

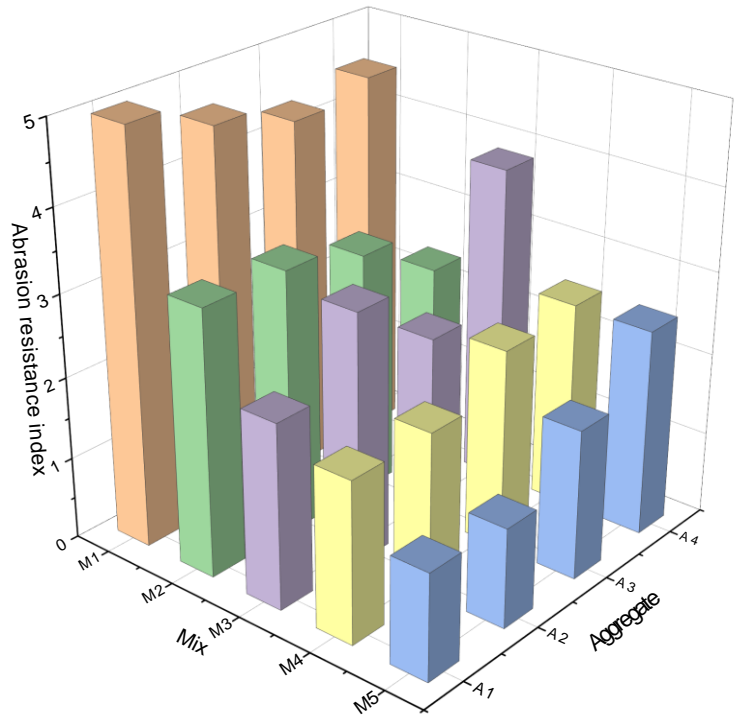

Fig. 17. Abrasion resistance index of the pervious concrete.

**3.7 Correlations among physical properties**

From the above results, it can be inferred that correlations exist in the porosity, permeability, and compressive strength of the specimens. Therefore, the data points of porosity-versus-permeability are shown in Fig. 18, where a power law was used to fit the data. It can be found the power law, $y = -0.379 + 0.054x^{0.99}$, gives an acceptable result for the fitting, which indeed approximates to a linear fit, since the power $0.99 \approx 1$. The fitting line also indicates when the porosity decreases to 7%, the permeability coefficient would approach 0, which means the specimen is impermeable. This agrees with reality. In the literature, linear (Akkaya & Çağatay, 2021; Cai et al., 2021; Shan et al., 2022), exponential (Anburuvel & Subramaniam; J. Huang et al., 2021; L. G. Li et al., 2021; Seifeddine et al., 2023) and also power law (Conzelmann et al., 2022; Kayhanian et al., 2012; Xu et al., 2018) had been used to correlate porosity with permeability coefficients of pervious concrete. Generally, the coefficients of fitting goodness $R^2$ in previous studies are higher than that in the present study. This is because the data set here is relatively large with many different mixes and aggregate sizes. These lead to fluctuations in the data and a reduction of $R^2$.

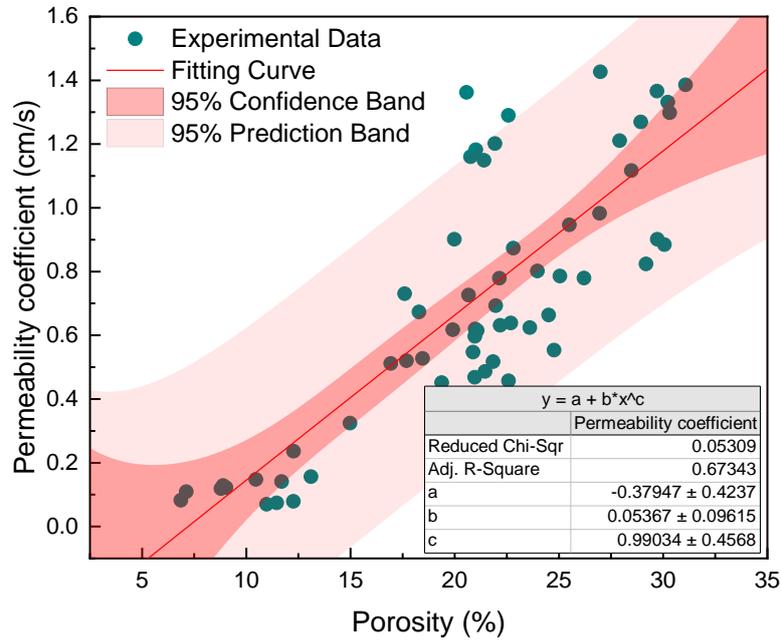

Fig. 18. Correlation between permeability coefficients and porosities.

For the correlation between porosity and compressive strength, an inverse trend exists in the data, which can be found in the illustration of Fig. 19. Similarly, a power law $y = 57.58 - 23x^{0.24}$ was used to fit the data. The coefficient of fitting goodness is greater than that of the correlation between porosity and permeability coefficient in Fig. 18. The power law also predicts when porosity approaches 0, the compressive strength approximates 57.58 MPa. This is also reasonable when considering only the macropores of the pervious concrete. In the literature, linear (Shen et al., 2021; Wang et al., 2022), logarithmic (L. G. Li et al., 2021), and exponential (Anburuvel & Subramaniam; Chindaprasirt et al., 2009) laws had been used to fit the correlation between porosity and compressive strength. The fitting results also depend on the sizes of data sets and mixes.

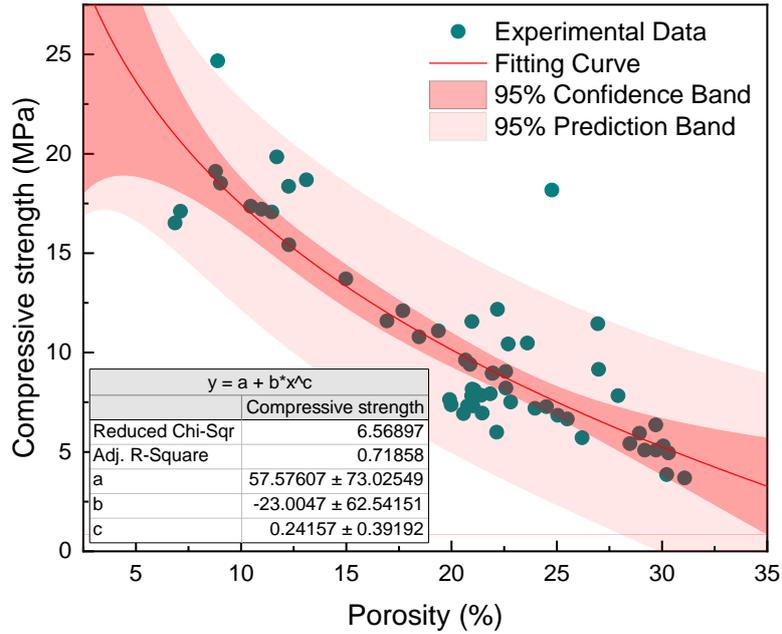

Fig. 19. Correlation between compressive strengths and porosities.

Additionally, the power law was also used to fit the correlation between the permeability coefficient and compressive strength of the pervious concrete. Results are shown in Fig. 20 where the fitting curve is $y = 2.62 - 0.833x^{0.37}$, which predicts the maximum permeability coefficient, that can be reached by the present mixes and aggregate sizes, would be 2.62 cm/s. This is a relatively high but reasonable value for the pervious concrete. In the literature, the fitting of correlation between the permeability coefficient and compressive strength is relatively rare, although linear fit (Yap et al., 2018) and exponential fit (Jike et al., 2022) had been used.

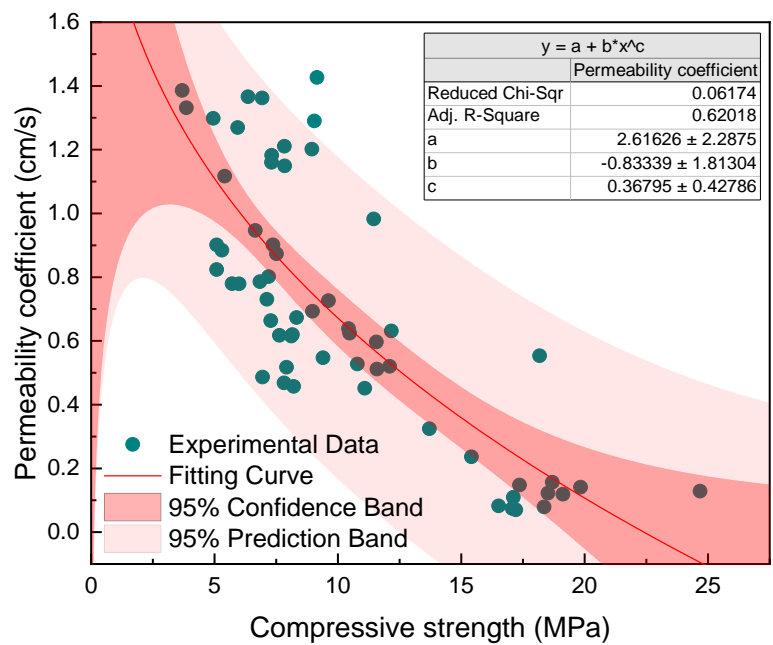

Fig. 20. Correlation between permeability coefficients and compressive strengths.

Moreover, the correlation between the abrasion resistance index and porosity was also studied. Results are shown in Fig. 21, where a linear law $y = 34.1 - 4.72x$ was used to fit the data. The coefficient of fitting goodness $R^2 = 0.723$, which is greater than those of other correlations in Figs. 18~20. The fitting line also predicts when the porosity approaches 0, the abrasion resistance index approximates 7.22 which is a theoretical maximum for the present mixes and aggregates. On the other hand, when the abrasion resistance index is reduced to 0, the porosity can theoretically reach 34.1%. These predictions can give a good understanding for the design of pervious concrete used as a pavement material. In the literature, few results on the correlation between the abrasion resistance index and porosity were reported.

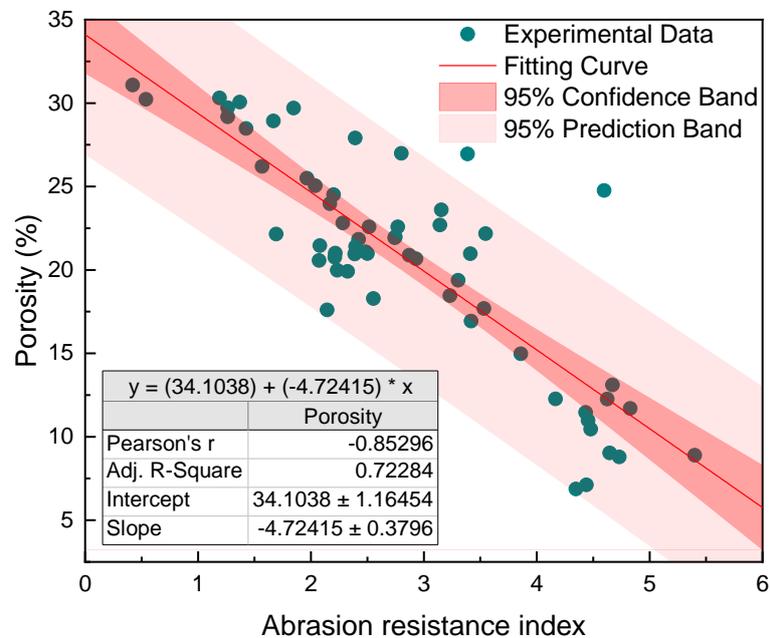

Fig. 21. Correlation between porosities and abrasion resistance indices.

**3.8 Comparison with previous studies**

The permeability coefficient and compressive strength of the pervious concrete are compared with those in previous studies (Aliabdo et al., 2018; El-Hassan et al., 2019; Wei Huang & Wang, 2022; Lima et al., 2022; A. Singh et al., 2022; Vieira et al., 2020; Yap et al., 2018). Results are shown in Figs. 22 and 23. Generally, the permeability coefficients and compressive strengths of the present pervious concrete are in the middle of the compared data. The maximum permeability coefficient in the present study is 1.645 cm/s (M5A2) which is in the third place among the compared data of the previous studies. The greatest maximum permeability coefficient appeared in

Ref. (Yap et al., 2018) with a value of 2.64 cm/s, which was achieved by using recycled aggregate to replace natural granite aggregate. However, the high permeability may reduce the mechanical strengths, which can be found in Fig. 23. Nevertheless, the specimens in Ref. (A. Singh et al., 2022) were with relatively high permeabilities and strengths, simultaneously. This is due to the multilayered form used in Ref. (A. Singh et al., 2022). On the other hand, the present minimum permeability coefficient with a value of 0.069 cm/s (M1A4) is the lowest among the compared data. This is attributed to the lower aggregate-to-cement ratio and higher water-to-cement ratio used in M1, which should be avoided in the mix design.

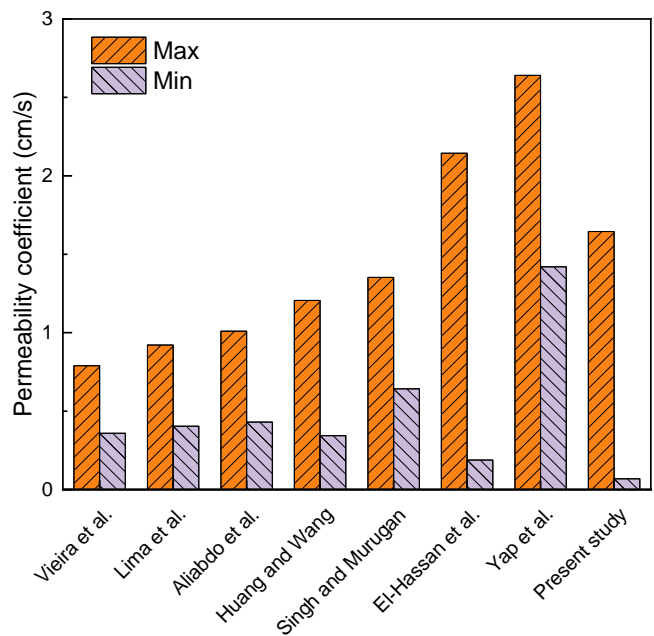

Fig. 22. Comparison of permeability coefficients of pervious concrete between previous and present studies.

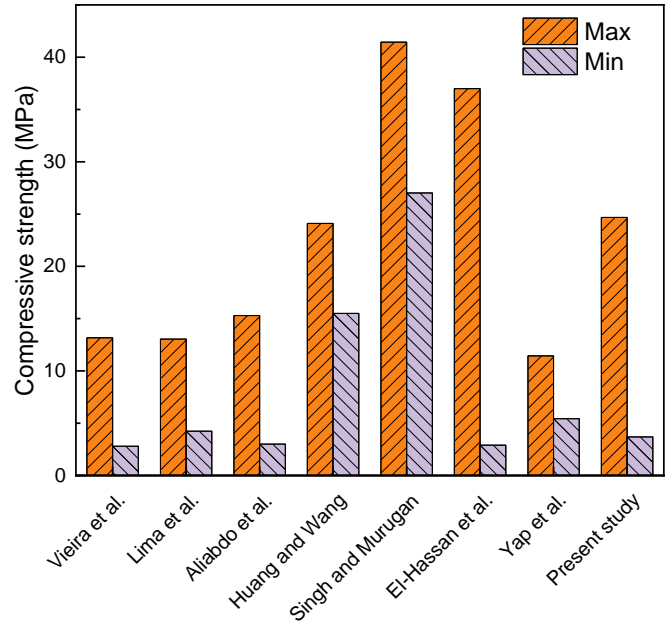

Fig. 23. Comparison of compressive strengths of pervious concrete between previous and present studies.

## 4 Conclusions

Pervious pavement design requires knowledge of the influences of aggregate sizes and mix proportions on the relevant properties and their correlations of pervious concrete, which were investigated in this work. Conclusions can be drawn as follows:

- In the manufacturing of pervious concrete, a lower aggregate-to-cement ratio led to redundant cement paste and porosity reduction, which may decrease the permeability coefficient but increase the compressive strength, thermal conductivity, and abrasion resistance of the pervious concrete. The aggregate-to-cement ratio between 3.5 and 4 gave relatively stable physical properties of the pervious concrete.

- The mix design of pervious concrete using a higher water-to-cement ratio increased the flowability of the cement paste, which may block the pores, especially those at the bottom, and then decrease the permeability coefficient but increase the compressive strength, thermal conductivity, and abrasion resistance of the pervious concrete. The water-to-cement ratio between 0.275 and 0.3 gave relatively better comprehensive physical properties of the pervious concrete.

- The aggregate sizes affected the physical properties of the pervious concrete relatively less in comparison with the mix proportions. In the same mix, the influence of aggregate sizes on the permeability coefficient was of more randomness and fluctuations than on the porosity, compressive strength, and abrasion resistance of the pervious concrete.

- According to the CT segmentation and reconstruction, the sizes of pores and cement paste in the pervious concrete were affected more by the aggregate sizes than by the mix proportions, whereas the position distribution of the pores and cement paste depended more on the mix proportions.

- The correlations among porosity, permeability coefficient, and compressive strength of the pervious concrete can be fitted by the power law, whereas that between the porosity and abrasion resistance index can be fitted by a liner law. The fitting parameters gave reasonable physical meanings for the pervious concrete.

## Acknowledgement

This work was supported by the National Natural Science Foundation of China (51908075).